# Inferring Political Alignments of Twitter Users

A case study on 2017 Turkish constitutional referendum

Kutlu Emre Yilmaz
Department of Computer Engineering
TOBB University of Economics and Technology
Ankara, Turkey
kyilmaz@etu.edu.tr

Osman Abul
Department of Computer Engineering
TOBB University of Economics and Technology
Ankara, Turkey
osmanabul@etu.edu.tr

*Abstract*—Increasing popularity of Twitter in politics is subject to commercial and academic interest. To fully exploit the merits of this platform, reaching target audience with desired political leanings is critical. This paper extends the research on inferring political orientations of Twitter users to the case of 2017 Turkish constitutional referendum. After constructing a targeted dataset of tweets, we explore several types of potential features to build accurate machine learning based predictive models. In our experiments, three-class support vector machine (SVM) classifier trained on semantic features achieves the best accuracy score of 89.9%. Moreover, an SVM classifier trained on full text features performs better than an SVM classifier trained on hashtags, with respective accuracy scores of 89.05% and 85.9%. Relatively high accuracy scores obtained by full text features may point to differences in language use, which deserves further research.

## I. INTRODUCTION

Social media have become an integral part of the political sphere, as increasing number of people resort to Twitter, a microblogging platform, for political news and information [1] or other political purposes [2]. How social media transforms the politics was evident in Obama's 2008 election success [3] and the rapid spread of Occupy protests [4].

Twitter receives commercial and academic interests with applications to political advertising, personalized news, product recommendations and research in political science. Reaching target population is crucial with any such an application which typically requires identifying various attributes of users. Since most of the attributes are not explicitly provided, the applications need to infer various attributes of the users by analyzing social media data they generate. The inferred attributes include age and gender [5], ethnicity [6], message impartiality [7] and political affiliation [8].

The subject of this study is to infer political alignments of Twitter users from their tweets. The case study selected is 2017 Turkish constitutional referendum which was held on April 16, 2017. The referendum was held for a controversial constitutional amendment, with voting options of *yes* to ratify or *no* to reject. Turkey is globally ranked 8th in the number of Twitter users [9]. Twitter has turned into a key media outlet for Turkish politics, after its organizing role in Gezi Park protests [10]. Similarly but not surprisingly, Twitter was subject to a battle between the supporters and the opponents of the constitutional change. The supporters of *yes* vote were mainly organized under #EVET (In Turkish: *yes*) hashtag, and opponents were mainly promoting #HAYIR (In Turkish: *no*) hashtag. We believe that Turkish referendum is an interesting case to extend the research on inferring political orientations.

This work infers the voting intentions of Turkish Twitter users during the referendum. Our contribution in this paper is threefold: First, we collected a targeted dataset of 2000 users with manually annotated political leanings of *yes*, *no* and *ambiguous*. Instead of hashtags or Twitter lists, we directly sample from trend topics and do not discard users with *ambiguous* political leanings. Second, we evaluate the effectiveness of four different types of linguistic features on a three-class classification task with class labels of *yes*, *no* and *ambiguous*. For this purpose, we report the performance of three different classifiers trained and tested on our dataset. The results show that support vector machine (SVM) trained on semantic features achieve the highest accuracy rate of 89.9%. Third, we analyze the relationship between classification accuracy and change in feature size. Experiments show that classification performance improves with increasing feature counts up to a certain point, and declines beyond that. The performance of all of the classifiers are compared with the performance of a baseline hashtag classifier we devise.

## II. RELATED WORK

Twitter is a popular microblogging platform, where users share text messages known as tweets. A retweet is a rebroadcast of an existing tweet. By retweeting, information and news spread rapidly [1]. A hashtag is a special token constructed by prepending a '#' character to a keyword. Hashtags are one word summaries of a tweet [11]. A trending topic is a hashtag, a word or a phrase that receives rapid increase in popularity. Trending topics are hot discussions, and are displayed on the left panel of the Twitter interface.

Large and diverse user population and public data of Twitter is subject to increasing academic interest. Particular type of research infers certain characteristics of Twitter users, including age, gender, regional origin [5], ethnicity [6], psychology [12] and language bias [7]. Among others, inferring users' political leanings [8, 13, 14] deserves attention with the increasing role of Twitter in politics [2].

Previous research study establishing effective features to build successful predictive models for inferring Twitter users' political leanings. Different features may encode networked [15], linguistic [8, 13], behavioral or profile [6] characteristics of users. Evaluating the effectiveness of profile information, tweeting behavior, linguistic content and social connections as features, the work [6] noted linguistic features as reliable and promising. The scope of this study is limited to linguistic features.

For linguistic features, the work [5] experiments with a complete set of unigrams and bigrams generated from data. An improved feature selection scheme evaluates the effectiveness of a specific subset of words (unigrams) and hashtags that are frequently used by manually annotated seed users [6]. To improve the discovery of relevant features, the work [8] start with seed hashtags and introduce a tag co-occurrence discovery procedure that scans the whole dataset for relevant hashtag features. Our study combines the feature selection procedure of [6] and [8].

In terms of linguistic feature types, [8] compile a list of hashtag features by selecting relevant hashtags that co-occur with a set of seed hashtags in a user's tweet. For the full text features, while [6] resorts to a variant of bag of words approach, [8] and [5] experiments with *tf-idf*, a special term weighting scheme widely used by information retrieval community [16]. Latent Semantic Analysis [8] and Latent Dirichlet Allocation (LDA) [6] are other experimented linguistic features that capture the richer semantic information in text. In a recent study, classifiers trained on unigram features achieve close performance to more complex feature types [15]. Also, deep learning and sentiment based linguistic features are evaluated in [13]. In this study, besides hashtags and bag-of-words, we also experiment with *tf-idf* and LDA features and compare results with select previous work.

To evaluate the effectiveness of proposed features, the literature report their classification performance on specially crafted datasets [5, 6, 8, 14]. SVMs [5, 8, 14] are the most popular classifiers [14] with accuracy rates up to 95%. Other reported classifiers are boosted decision trees [6], regression [13] and various estimation methods [17, 18]. In this study SVM, Decision Tree and Random Forests are employed.

One caveat is the validity of constructed datasets. Datasets are biased and not all Twitter users clearly express their political opinions in their messages [14]. The experiments in [14] with different datasets in varying levels of difficulty, report previous results as optimistic in their reported levels of accuracy. However, leaning solely on hashtags for constructing different datasets still introduces some form of bias [14]. In this study, we construct a novel dataset by directly sampling data from trend topics. The advantages of this approach in reducing bias are two-fold: First, since a trend topic might be a word, a phrase or a hashtag, our focus is not only limited to hashtags. Second, trend topics receive more attention than hashtags.

Previous datasets are tailored for classification tasks with clearly separated political leanings (e.g. Democrats and Republicans). [13] is an exception with seven scales of political leanings. However, still, on their study too, users with undefined political leanings are labeled as *ambiguous*, and are usually discarded during dataset construction. However, real world Twitter activity is contaminated with spam and noise, naturally requiring a reconsideration of the *ambiguous* user orientations. Thus, contrary to the previously created minimum noise datasets, we introduce a novel dataset with three-class labels of *yes*, *no*, *ambiguous*. We report the results of a three-class classification task, rather than conventional practices of two.

In international context, Fang et al. [19] study Scottish independence referendum and propose a topic based Naive Bayesian classifier which performs better than baseline Naive Bayesian classifier. Barbera [17] use ideal point estimation method to infer political positions of Twitter users in Germany, Spain and United States, and validate proposed methodology by comparing prediction results with actual voter records in U.S. The subject of this study is 2017 Turkish constitutional referendum which was held on April 16, 2017.

### III. DATA

Data collection was performed using Twitter Streaming API, and spans over the time period between March 21st and May 14th, namely, four weeks before and after the referendum date of April 16, 2017. Streaming API provides real time access to 1% of the entire stream of public tweets which could be filtered by the presence of certain keywords to be tracked or the geographic locations of users. We filter by trend topic keywords and user locations in Turkey. Every 10 minutes, we query Twitter API and obtain a list of trend topics. We manually check the consolidated list of trend topics at least two times a day, and add the politically relevant ones to the list of keywords to track. During the period of 8 weeks, a total of 1871 unique politically relevant trend topics were tracked, resulting in approximately 33 million tweets collected. The full dataset contains approximately 22 million retweets produced by 1.4 million users, of which 311,317 retweeted 10 or more times. The aim of data collection was a broad analysis of the different aspects of the referendum period. Therefore, in this study, we only use a small subset of the full dataset.

For the purpose of this study, a dataset of 2000 users are randomly selected from the population of users who retweet 10 more times. We apply a stratified sampling procedure, and observe a close match between the mean, median and quartiles of the full and sampled dataset. The political alignment of each user is labelled as either *yes*, *no* or *ambiguous* by manually examining the content of their tweets. Political leanings of *yes* and *no* indicate the voting options in the referendum. Whereas, the label *ambiguous* represents, accounts which produce spam, politically irrelevant material, or users whose political orientation cannot be determined. After this procedure, 826 users were annotated as voters of *yes*, 782 users as voters of *no*, and the remaining 392 users were assigned to the class of *ambiguous*. We manually checked the users marked as *ambiguous*, and the majority of them are either bots or spammers that aim to reach a broader audience by deliberately injecting politically relevant material in their messages. This leads to an increase in the number of *ambiguous* cases.

## IV. FEATURES

In this section, we describe several linguistic features used in this study. The term *document* denotes the set of all tweets produced by a particular user.

### A. Hashtags

Hashtags are special tokens that serve as topic markers which allows for users to easily follow ongoing interactions [11]. Table I gives the frequency of the top 10 hashtags in the full dataset. Unsurprisingly, the most popular hashtags are #EVET (*yes*) and #HAYIR (*no*) which represent the two voting options in the referendum. Hashtags have a similar role to the role of abstracts in documents [8]. Thus, they are important features summarize message content.

TABLE I. TOP-10 MOST OCCURRING HASHTAGS IN THE FULL DATASET

| |
| --- |
| #EVET (*yes*, 872172); #HAYIR (*no*, 746272); #Hayır (*no*, 698258); #Evet (*yes*, 448342); #hayır (*no*, 286996); #BugünHayırÇıkacak (239689); #HayırDahaBitmedi (232863); #izmirescort (214700); #YarınHayırÇıkacak (193486); #EVETdeTarihYaz (168533) |

a. English translation are given in italics. Only translations of common words are provided.

For the features setup, our approach is similar to [6]. However, the scope of their hashtag discovery procedure is limited to the preselected seed users. To fix this, we also apply a tag co-occurrence discovery procedure proposed by [8]. Briefly, we construct two separate ranked lists of hashtag features, one for the voters of *yes* and the other for the voters of *no*. Then, we select top-$k$ ranked items from each list and consolidate them into a single feature set of size $2 \times k$. To understand the impact of the feature size on classification accuracy, we experiment with top-$k$ feature subsets for different values of $k$. Table II lists experimented $k$ values.

TABLE II. EXPERIMENTED FEATURE TYPES AND SIZES

| Feature Type | Experimented Feature Subset Sizes (k) |
| --- | --- |
| Hashtag | 5, 10, 25, 50, 100, 250, 500, 1K, 2.5K, 5K, 7.5K * |
| Bag of Words | 5, 50, 100, 250, 500, 1K, 2.5K, 5K, 10K, 20K, 25K * |
| TF-IDF | 100, 250, 500, 1K, 2.5K, 5K, 10K, 20K, 50K |
| LDA | 10, 20, 50, 100, 200, 500, 1K, 2.5K, 5K, 10K, 15K |

b. The capital K denotes 1000 (e.g. 2.5K is 2500). * Final feature set size is two times k.

A detailed description of feature selection procedure is as follows. Two separate sets of seed hashtags, $S_{YES}$ and $S_{NO}$, are initialized as,

$$S_{YES} = \{ \#EVET, \#Evet, \#evet \} \quad S_{NO} = \{ \#HAYIR, \#Hayır, \#hayır \}$$

Let H be the set of tweets that contains a hashtag h. Then, we calculate Jaccard similarity score, $J(S_{YES}, h)$, for each hashtag h that co-occurs with a seed hashtag $h_s$ in a tweet, where $h_s \in S_{YES}$. To calculate the similarity score $J(S_{NO}, h)$, the same procedure is repeated for hashtag h and the set $S_{NO}$, where,

$$J(S_{YES}, h) = \frac{|S_{YES} \cap H|}{|S_{YES} \cup H|} \quad , \quad J(S_{NO}, h) = \frac{|S_{NO} \cap H|}{|S_{NO} \cup H|}$$

Finally, we assign each hashtag h to the feature list for which it attains its maximum Jaccard similarity score. This process yields a total of 8804 hashtags associated with the supporters of *yes* vote, and 8295 with the *no* vote.

Table III shows top-10 items of each feature list and corresponding similarity scores. We observe the highest similarity scores between #EVET and #HAYIR hashtags, indicating that users heavily prefer to use supporting and opposing hashtags together in their messages. A similar type of behavior, known as *content injection*, was reported in [8]. However, [8] do not provide any quantitative information about the level of content injection, so it is not clear whether it is as significant as in the case of the Turkish referendum.

TABLE III. TOP-10 HASHTAGS THAT CO-OCCUR WITH SEED HASHTAGS AND RESPECTIVE SIMILARITY SCORES

| #EVET (*yes*) | #HAYIR (*no*) |
| --- | --- |
| #HAYIR (0.2335) | #EVET (0.3341) |
| #16Nisan (0.0131) | #YarınHayırÇıkacak (0.0077) |
| #BenimKararımNet (0.0118) | #BugünHayırÇıkacak (0.0064) |
| #EVETdeTarihYaz (0.0116) | #NedenHAYIRÇünkü (0.0060) |
| #Referandum (0.0077) | #HayırKazandıYSKÇaldı (0.0054) |
| #16NisanYükseliş (0.0073) | #Referandum2017 (0.0051) |
| #GüçlenenTürkiyeİçin (0.0071) | #HayırDahaBitmedi (0.0049) |
| #TekYürekEVET (0.0064) | #BahçeliHayırDiyor (0.0046) |
| #Türkiye (0.0054) | #CHP (0.0042) |
| #EvetGelecektir (0.0048) | #SokaktaHayırVar (0.0040) |

### B. Bag of Words

In bag-of-words model each document is represented as a sparse vector over all unique terms (words) in a corpus. For each term, the corresponding element of the document vector is defined as the number of times that respective term appears in the document.

The proposed feature selection methodology is as follows. After preprocessing text by removing hashtags, mentions and URLs, we start with two separate sets of seed hashtags, $S_{YES}$ and $S_{NO}$. Then, we repeat the same feature selection procedure previously applied for a hashtag h, but this time for a word w that co-occurs with a seed hashtag in a tweet. Finally, we create two separate ranked lists of word features, and experiment with top-$k$ feature subsets for different values of $k$ (Table II).

### C. TF-IDF

In information retrieval, *tf-idf* is a special term weighing scheme that determines the relative importance of a word (term) in a document [16]. Terms unique to a document are rewarded in proportion to the number of times they appear in that document (term frequency - TF), while terms that frequently appear in other documents are penalized in proportion to the frequency of appearance in other documents (inverse document frequency - IDF). After preprocessing text by removing hashtags, mentions and URLs, we obtained a total number of 297,000 terms in the corpus. For the *tf-idf* features, we only consider top-k most frequent terms in the corpus for different values of k (Table II).

*D. Latent Dirichlet Allocation*

Latent Dirichlet Allocation (LDA) is a generative probabilistic model proposed for automatically inferring topics over a set of documents [20]. Each document is represented as a multinomial distribution over topics, and each topic is represented as a multinomial distribution over words with the parameters of the multinomial distributions. The aim is to maximize the probability of each document to contain certain topics, and the probability of each word to represent a certain topic. The number of topics is a preset parameter. The output is a ranked list of word distribution for each topic.

For the text preprocessing, we kept hashtags and removed mentions and URLs. As the LDA implementation we use Gensim, an open source Python library designed for unsupervised text modelling tasks [21]. The model is trained on a sample of 100,000 randomly selected documents for 400 iterations, with a choice of topic number as 10 and dictionary size of 100,000. The motivation behind choosing 10 as the number of topics is that the number should be small enough to keep politically relevant topics intact, while being large enough to separate nonpolitical topics such as advertisements, spam and noise. In theory, LDA should produce more descriptive features by incorporating semantic information hidden in the text and thus should improve classification accuracy.

Table IV lists top-10 terms for nonpolitical topics derived by the LDA. As hypothesized, T5, T7 and T8 are highly discriminative nonpolitical topics, and features derived from them should increase prediction accuracy for ambiguous political leanings. T5 is mostly about entertainment, T7 is about sports and football, T8 is about advertising of sex services. Final feature set is constructed by selecting top-k terms from the distributions of words produced for each topic, and consolidating them into a single feature set of size at most 10xk for different values of k (Table II).

TABLE IV. WORD DISTRIBUTIONS FOR NONPOLITICAL LDA TOPICS

| Topic | Top-10 Word Distribution |
|---|---|
| T5 | ücretsiz (*free*, 0.042); indir (*download*, 0.042); play (0.042); google (0.042); oy (*vote*, 0.018); takip (*follow*, 0.017); rt (0.016); fav (0.015); etmeyenler (0.015); geçersizdir (0.015) |
| T7 | son (*end*, 0.016); rt (0.014); nisan (*april*, 0.014); anket (*survey*, 0.012); #survivor (0.010); evet (*yes*, 0.008); ol (0.008); kazan (*win*, 0.007); oyun (*game*, 0.006); maç (*football match*, 0.005) |
| T8 | #izmirescort (0.070); yaşında (*age*, 0.059); bornova (0.033); 21 (0.032); gerçek (*real*, 0.032); izmir (0.032); #bornovaescort (0.031); 0539 (0.031); 1820 (0.031); 914 (0.031); kendi (0.030) |

c. English translation are given in italics. Only translations of common words are provided.

V. EXPERIMENTS AND EVALUATION

To investigate the effectiveness of different types of features for the political orientation prediction problem, we experiment with three classifiers: SVM; Random Forests; Decision Tress. For this purpose, we evaluate the performance of classifiers trained on features (Hashtag, Bag of Words, TF-IDF, and LDA) of different sizes (Table II). Obtained results are compared to the baseline performance of a simple hashtag based classifier. For three-class (*yes*, *no*, *ambiguous*) classification task, we report performance measures of accuracy and confusion matrix obtained through 10-fold cross validation. We use Weka [22] as the toolbox.

For the baseline classification task, the devised scheme is a simple three-class majority vote classifier. In this scheme, each occurrence of #EVET hashtag in a user's tweets, counts as a vote for the class label of *yes*. Similarly, each occurrence of #HAYIR hashtag is a vote for the class label *no*. The user is assigned to the class label that receives the highest vote. In the case of a tie, the user's orientation is labelled as *ambiguous*.

In political orientation prediction literature, SVMs are the most popular classifiers [14]. For the case of tree-based learners, only one study uses boosted decision trees [6]. We experimented with both Gaussian and linear kernel SVMs. The performance of linear SVM was slightly better. Random forests and C4.5 [24] are the tree-based classifiers trained and tested on our dataset. SVM [23] fits a non-probabilistic model from the training data by separating instances into different classes with maximum margin between the decision boundaries. Random forests are ensemble of decision trees. The final decision for an instance is the majority vote class label.

In line to previous studies, e.g. [14], SVMs are the best performing classifiers in our three-class classification tasks. As shown in Table V, the best overall classification accuracy of 89.9% is achieved with a linear SVM trained on LDA features. As reported in Fig. 1d, random forests and C4.5 classifiers trained on our dataset achieve a maximum accuracy score of 88.05% and 81%, respectively. In Fig. 1, for all feature types, random forests achieve close performance to the SVM, and decision trees are outperformed by both classifiers. The best accuracy score for the baseline classifier is 75.6% (Table V).

Figure 1 shows change in classification performance with feature size. For all feature types, an increase in feature size is followed by a rise in classification accuracy up to a certain point. After reaching the peak accuracy value, additional features only degrade the performance. Probably, the noise introduced by additional features that are irrelevant or less discriminative lead to a decline in classification performance. As decision trees are more susceptible to overfitting due to noise and irrelevant features, C4.5 is the most unstable classifier against increasing feature size. As succinct expressions of message content, hashtags are highly representative and balanced features. Compared to other features, classifiers trained on hashtags are less sensitive to changes in feature size (Fig. 1a). A drawback of this is the least performance gain with increasing feature size and hashtags achieve the lowest maximum accuracy score over all other feature types.

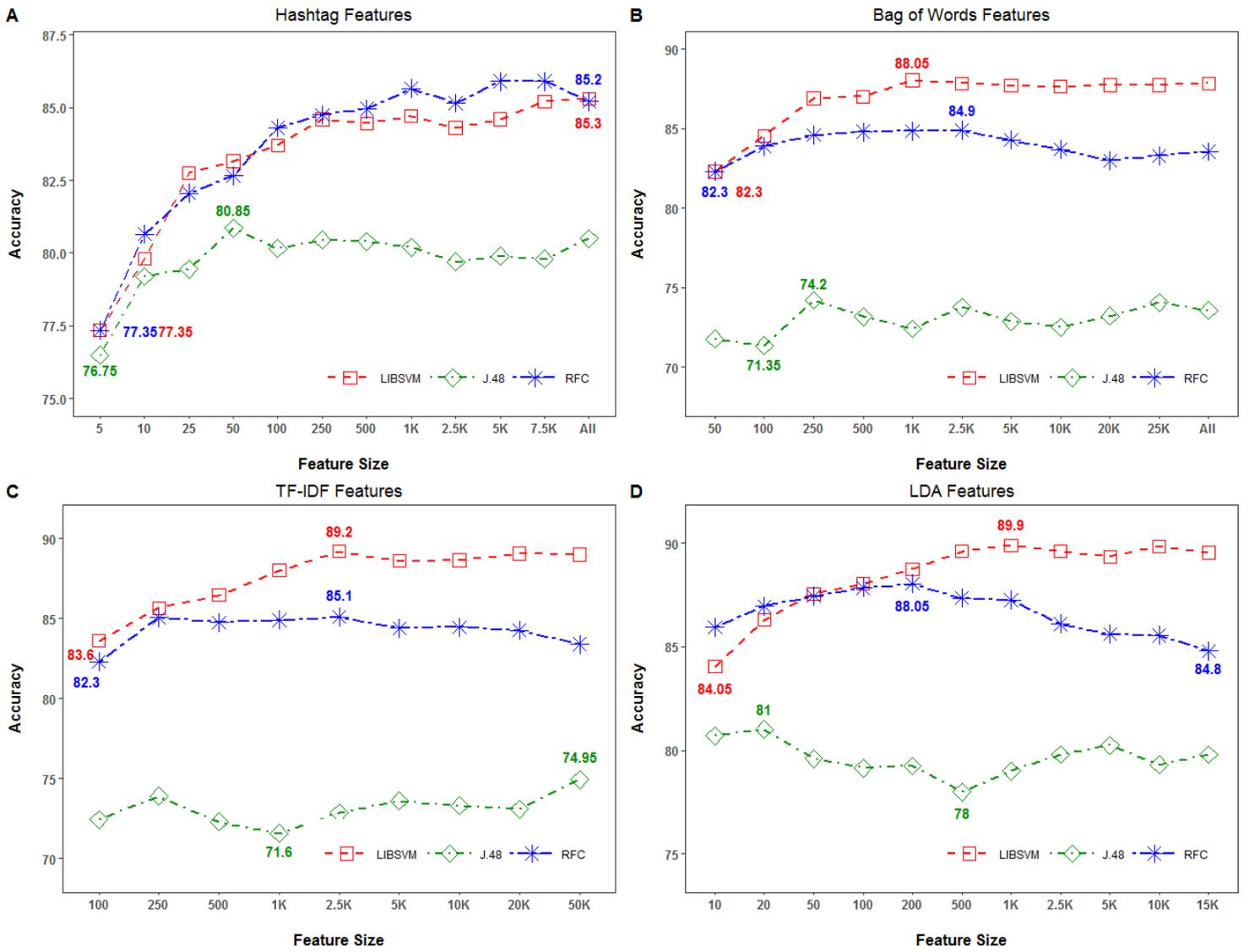

For three classifiers, the change of accuracy with different values of k and feature types: A) Hashtags B) Bag of Words C) TF-IDF D) LDA. For each classifier, only the maximum and minimum accuracy scores that classifier achieves is explicitly annotated. RFC denotes Random Forests.

TABLE V. CONFUSION MATRICES AND ACCURACY VALUES FOR DIFFERENT FEATURES

| Feature Type | Confusion Matrix | Accuracy | Classifier |
|---|---|---|---|
| Hashtag | $\begin{matrix}(yes) & 599 & 15 & 212 \\ (no) & 10 & 591 & 181 \\ (amb.) & 41 & 29 & 322\end{matrix}$ | 75.6% | Baseline |
| Hashtag | $\begin{matrix}yes & 751 & 30 & 45 \\ no & 25 & 699 & 58 \\ amb. & 80 & 44 & 268\end{matrix}$ | 85.9% | RF |
| Bag of Words | $\begin{matrix}yes & 766 & 30 & 30 \\ no & 25 & 728 & 29 \\ amb. & 74 & 51 & 267\end{matrix}$ | 88.05% | SVM |
| TFIDF | $\begin{matrix}yes & 780 & 19 & 27 \\ no & 22 & 736 & 24 \\ amb. & 66 & 61 & 265\end{matrix}$ | 89.05% | SVM |
| LDA | $\begin{matrix}yes & 783 & 17 & 26 \\ no & 19 & 738 & 25 \\ amb. & 70 & 45 & 277\end{matrix}$ | 89.9% | SVM |

Rao et al. [5] report *tf-idf* performing worse than bag-of-words features. We observe a converse trend for the optimum accuracy scores. In Fig. 1, SVM trained on bag-of-words (88.05%) and *tf-idf* (89.05%) features perform better than hashtag features (85.9%), which is also the case in [25]. Whereas, for tree-based classifiers, hashtags perform better than bag-of-words and *tf-idf* features, confirming the results in [8]. As shown in Table V, for users with political leanings of *yes* and *no*, classifiers incorporating bag-of-words, tf-idf and LDA features perform better than the ones incorporating hashtag features. The main reason behind this is the methodological differences in dataset preparation [14]. Contrary to the deterministic approaches (e.g. hashtags, Twitter lists) used in dataset construction, our methodology relies on directly sampling data from trend topics. Thus our dataset is noisier and less biased towards politically active users with profound hashtag use. Therefore, LDA and full text features capture more information than hashtags, at the expense of introducing additional noise which could be better handled by SVMs compared to tree-based classifiers.

SVM and random forest classifiers trained on LDA features achieve the highest accuracy values, which is the case for C4.5 trained on hashtag features. Moreover, compared to other feature types, C4.5 trained on bag-of-words and tf-idf features, suffer from more than 5% loss in accuracy. Consequently, optimum accuracy scores for C4.5 trained on bag-of-words (74.2%) and tf-idf (74.95%) features are worse than the baseline classifier (75.6%), while C4.5 performs better when trained on hashtag (80.85%) and LDA (81%) features. As decision trees are more susceptible to overfitting, their performance improves when trained on LDA and hashtag features which consider textual semantics and are thus more descriptive.

For SVM classifiers trained on LDA features, a particular improvement is observed in the number of correctly classified instances for the class *ambiguous* (Table V). As hypothesized, compared to others, LDA features seem to better discriminate politically irrelevant topics such as advertisements and spam, for which, other feature types have difficulty in capturing. An interesting fact is the substantial performance achievement of baseline classifier by outperforming other classifiers with more than %15 improvement in prediction accuracy for *ambiguous* user orientations (Table V). A close inspection of the dataset reveals the reason behind. With different goals, both politically active users and spammers follow a pattern of multiple hashtag use to reach a wider audience by including *yes* and *no* leaning hashtags together in their tweets.

## VI. CONCLUSIONS

In this paper, we studied political orientation inference problem for the case of 2017 Turkish constitutional referendum. We studied the effectiveness of different types of feature sets by evaluating the performance of classifiers trained and tested on our dataset.

For this purpose, we constructed a dataset with three class labels of *yes*, *no*, and *ambiguous*. Contrary to binary classification tasks in previous works, we introduced a third class label of *ambiguous* for users whose political leanings cannot be determined. Our goal was to simulate real world classification tasks in which irrelevant items and noise also has to be detected automatically. The best accuracy score of 89.9% is achieved with a SVM trained on features that incorporate semantic information. Also, increase in feature set size improves classification accuracy up to a certain point, after which performance declines.

In general, our findings are in line with previous research on political orientation prediction. However, on several issues, our findings diverge with previous work. Contrary to previous research, full text features perform remarkably better than hashtag features. This points to a difference in language use between yes and no voters, a topic that deserves further investigation. Moreover, more than 75% classification accuracy of the baseline classifier with just two seed hashtags indicates to a polarized user mass and deserves further research as well.